\documentclass[twocolumn]{aastex631}

\usepackage[T1]{fontenc}
\usepackage{physics}

\begin{document}

\title[IMBH migration \& accretion]{ From Seeds to Supermassive Black Holes: Capture, Growth, Migration, and Pairing in Dense Proto-Bulge Environments}

\correspondingauthor{Yanlong Shi}
\email{yanlong.astro@outlook.com}

\newcommand{\caltech}{TAPIR, MC 350-17, California Institute of Technology, Pasadena, CA 91125, USA}
\newcommand{\cita}{Canadian Institute for Theoretical Astrophysics, University of Toronto, Toronto, ON M5S 3H8, Canada}

\author[0000-0002-0087-3237]{Yanlong Shi}
\affiliation{\caltech}
\affiliation{\cita}

\author[0000-0002-4086-3180]{Kyle Kremer}
\affiliation{\caltech}

\author[0000-0003-3729-1684]{Philip F. Hopkins}
\affiliation{\caltech}

%\collaboration{20}{(AAS Journals Data Editors)}



\begin{abstract}

The origins and mergers of supermassive black holes (BHs) remain a mystery. We describe a scenario from a novel multi-physics simulation featuring rapid ($\lesssim 1\,$Myr) hyper-Eddington gas capture by a $\sim 1000\,{\rm M}_{\odot}$ ``seed'' BH up to supermassive ($\gtrsim 10^{6}\,\rm M_{\odot}$) masses, in a massive, dense molecular cloud complex typical of high-redshift starbursts. Due to the high cloud density, stellar feedback is inefficient and most of the gas turns into stars in star clusters which rapidly merge hierarchically, creating deep potential wells. Relatively low-mass BH seeds at random positions can be ``captured'' by merging sub-clusters and migrate to the center in $\sim1$ free-fall time (vastly faster than dynamical friction). This also efficiently produces a paired BH binary with $\sim 0.1$\,pc separation. The centrally-concentrated stellar density profile (akin to a ``proto-bulge'') allows the cluster as a whole to capture and retain gas and build up a large (pc-scale) circum-binary accretion disk with gas coherently funnelled to the central BH (even when the BH radius of influence is small). The disk is  ``hyper-magnetized'' and ``flux-frozen'': dominated by a toroidal magnetic field with plasma $\beta \sim 10^{-3}$, with the fields amplified by flux-freezing. This drives hyper-Eddington inflow rates $\gtrsim 1\,\rm M_\odot yr^{-1}$, which also drive the two BHs to nearly-equal masses. The late-stage system appears remarkably similar to recently-observed high-redshift ``little red dots.'' This scenario can provide an explanation for rapid SMBH formation, growth and mergers in high-redshift galaxies.

\end{abstract}



\section{Introduction}
\label{sec:intro}

Observations of high-redshift ($z\gtrsim 7$, $\sim 0.7\,\rm Gyr$ since the Big Bang) quasars demonstrate the existence of supermassive black holes (SMBHs) of $>10^9\,\rm M_\odot$ at these very early stages of the Universe \citep{FanNarayananLupton_2001AJ....122.2833F,YangWangFan_2020ApJ...897L..14Y,WangYangFan_2021ApJ...907L...1W}. How supermassive BHs form at all -- let alone how they grow so quickly -- remains a major unsolved problem \citep{InayoshiVisbalHaiman_2020ARA&A..58...27I,VolonteriHabouzitColpi_2021NatRP...3..732V}. Evidence from low-redshift studies argues such SMBHs grow from ``seed'' masses (perhaps even as low as stellar $10-100\,{\rm M}_{\odot}$) via gas accretion \citep{YuTremaine_2002MNRAS.335..965Y}. But assuming accretion is radiatively inefficient (and spherical), the Eddington limit is commonly cited as setting a classical upper limit to the BH accretion rate, which is defined as $\dot{M}_{\rm Edd}=M_{\rm BH}/(\epsilon_{\rm ref} t_{\rm Sal})$, where $t_{\rm Sal} = \kappa_{\rm es} c/(4\pi G) \approx 45\,\rm Myr$ is the Salpeter time and $\epsilon_{\rm ref}$ is an assumed radiative efficiency \citep[typically 0.1;][]{InayoshiVisbalHaiman_2020ARA&A..58...27I}. Thus, to grow to SMBH masses from ``seed'' masses $\ll 10^{6} \rm M_\odot$ \citep{GreeneStraderHo_2020ARA&A..58..257G} at high redshifts generally requires {either} accretion well above this naive Eddington limit \citep{InayoshiHaimanOstriker_2016MNRAS.459.3738I,ShiKremerGrudic_2023MNRAS.518.3606S}, {or} exotic IMBH seed formation scenarios like the direct collapse of un-fragmented giant molecular clouds to single hyper-massive stars \citep[with SMBH masses themselves;][]{BrommLoeb_2003ApJ...596...34B}, exotic dark matter/new physics processes \citep{XiaoShenHopkins_2021JCAP...07..039X}, or runaway mergers of stars \citep{PortegiesZwartBaumgardtHut_2004Natur.428..724P,ShiGrudicHopkins_2021MNRAS.505.2753S,RantalaNaabLahen_2024MNRAS.531.3770R}.

Theoretical work on small-scale ($\sim $\,au or horizon-scale) accretion physics has shown that super-Eddington accretion 
%\kyle{(defined as $\dot{M}_{\rm Edd}=M_{\rm BH}/(\epsilon_{\rm ref} t_{\rm Sal})$, where $t_{\rm Sal} = \kappa_{\rm es} c/(4\pi G) \approx 0.45\,\rm Gyr$ and $\epsilon_{\rm ref}$ is assumed to be 0.1)} 
is possible \citep{Begelman_1979MNRAS.187..237B,BlandfordBegelman_2004MNRAS.349...68B,InayoshiHaimanOstriker_2016MNRAS.459.3738I}. This is primarily enabled by the ``photon trapping'' effect, in which photons are trapped in the strong accretion flow ($\sim 1000\dot{M}_{\rm Edd}$) and advected into the BH (radiating inefficiently). This is also demonstrated by numerical simulations \citep{JiangStoneDavis_2014ApJ...796..106J,SadowskiNarayanTchekhovskoy_2015MNRAS.447...49S,JiangStoneDavis_2019ApJ...880...67J} where a sustainable phase of super-Eddington accretion is observed. But these simulations focus only on the accretion disk scales well interior to radii where the BH dominates the potential: another (perhaps more challenging) requirement for super-Eddington accretion is that the BH can gravitationally capture sufficient gas (from its turbulent, star-forming ISM environment) to sustain such accretion for the time needed to grow. This could be especially challenging in high-redshift galaxies where star formation is clumpy and bursty, potentially scattering BH seeds and inhibiting large-scale, coherent accretion flows \citep{MaHopkinsMa_2021MNRAS.508.1973M,ByrneFaucher-GiguereStern_2023MNRAS.520..722B}. Furthermore, classical accretion disk models \citep{ShakuraSunyaev_1973A&A....24..337S} predict such a super-Eddington accretion disk should be violently gravitationally unstable and fragment outside of tens of gravitational radii ($G M_{\rm BH}/c^{2}$) from the horizon \citep{Goodman_2003MNRAS.339..937G}.

Some of the problems above are explored in \citet{ShiKremerGrudic_2023MNRAS.518.3606S} where we embedded BH seeds into star-forming giant molecular clouds (GMCs). We found that significant BH accretion can ``stochastically'' occur if a dense clump (pc or sub-pc scale) formed via turbulence and stellar feedback shocks \citep{Klessen_2000ApJ...535..869K,MacLowKlessen_2004RvMP...76..125M,McKeeOstriker_2007ARA&A..45..565M} is {sufficiently} gravitationally bound and close (sub-pc) to the BH, and if the gas is sufficiently dense and cold \citep{InayoshiHaimanOstriker_2016MNRAS.459.3738I}. In practice, this ``clump accretion'' only occurs if both the gas and the BH dynamics are dominated by the GMC's self-gravity, and stellar feedback is globally inefficient (unable to rapidly unbind all the gas in the GMC as soon as the first massive stars form; \citealt{ShiKremerGrudic_2023MNRAS.518.3606S}), which is true for GMCs with high initial surface density \citep{GrudicHopkinsFaucher-Giguere_2018MNRAS.475.3511G,GrudicGuszejnovHopkins_2021MNRAS.506.2199G,ChevanceKrumholzMcLeod_2023ASPC..534....1C}. In \citet{ShiKremerHopkins_2024arXiv240512164S} we extended these to include physically-motivated feedback from BH accretion in the forms of radiation \citep{JiangStoneDavis_2019ApJ...880...67J}, accretion-disk winds and jets \citep{SilkRees_1998A&A...331L...1S,BlandfordBegelman_1999MNRAS.303L...1B}, and cosmic rays \citep{GuoOh_2008MNRAS.384..251G,GuoMathews_2012ApJ...756..181G,Zweibel_2017PhPl...24e5402Z}. We showed that for reasonable feedback parameters, BHs can still grow rapidly and their growth is primarily ``fueling-limited'' until they reach supermassive masses and the ``feedback-limited'' regime. But this alone does not address many of the key questions above. 

Meanwhile, in a series of papers using a different numerical method, physics, and technical approach, \citet{HopkinsGrudicSu_2024OJAp....7E..18H,HopkinsSquireSu_2024OJAp....7E..19H} followed gas accreting onto a pre-existing supermassive BH from cosmological initial conditions including star formation, and found a novel type of BH accretion disk: one which is ``hyper-magnetized'' (ratio of thermal-to-magnetic pressure $\beta \ll 10^{-2}$) and ``flux-frozen'' (dominated by a primarily toroidal magnetic field amplified from typical ISM fields by flux-freezing). In these papers and a subsequent analytic study \citep{HopkinsSquireQuataert_2024OJAp....7E..20H}, they showed that these disks could resolve a number of problems posed by classic accretion disk models (like \citealt{ShakuraSunyaev_1973A&A....24..337S}) which assume $\beta \gg 1$, most notably that the disk can remain stable (with Toomre $Q \gg 1$) out to $\gtrsim$\,pc scales even with accretion rates thousands of times larger than the nominal Eddington limit.

Here, we present a case study of one of the massive, high-density cloud-complex simulations in \citet{ShiKremerGrudic_2023MNRAS.518.3606S,ShiKremerHopkins_2024arXiv240512164S}, representative of plausible starburst conditions in high-redshift progenitors of massive galaxies. We show that this demonstrates a number of remarkable phenomena that could unify many previously-proposed mechanisms to enable rapid formation of a well-defined dynamical center, ``trapping'' of BHs, efficient and rapid migration of BH pairs to the center, efficient gas capture, and hyper-Eddington accretion through a hyper-magnetized accretion disk. We show this is able to take a BH from ``seed'' masses to truly supermassive masses in less than one Myr, providing a natural theoretical scenario for high-redshift BH growth.

\begin{figure*}
    \centering
    \includegraphics[width=\linewidth]{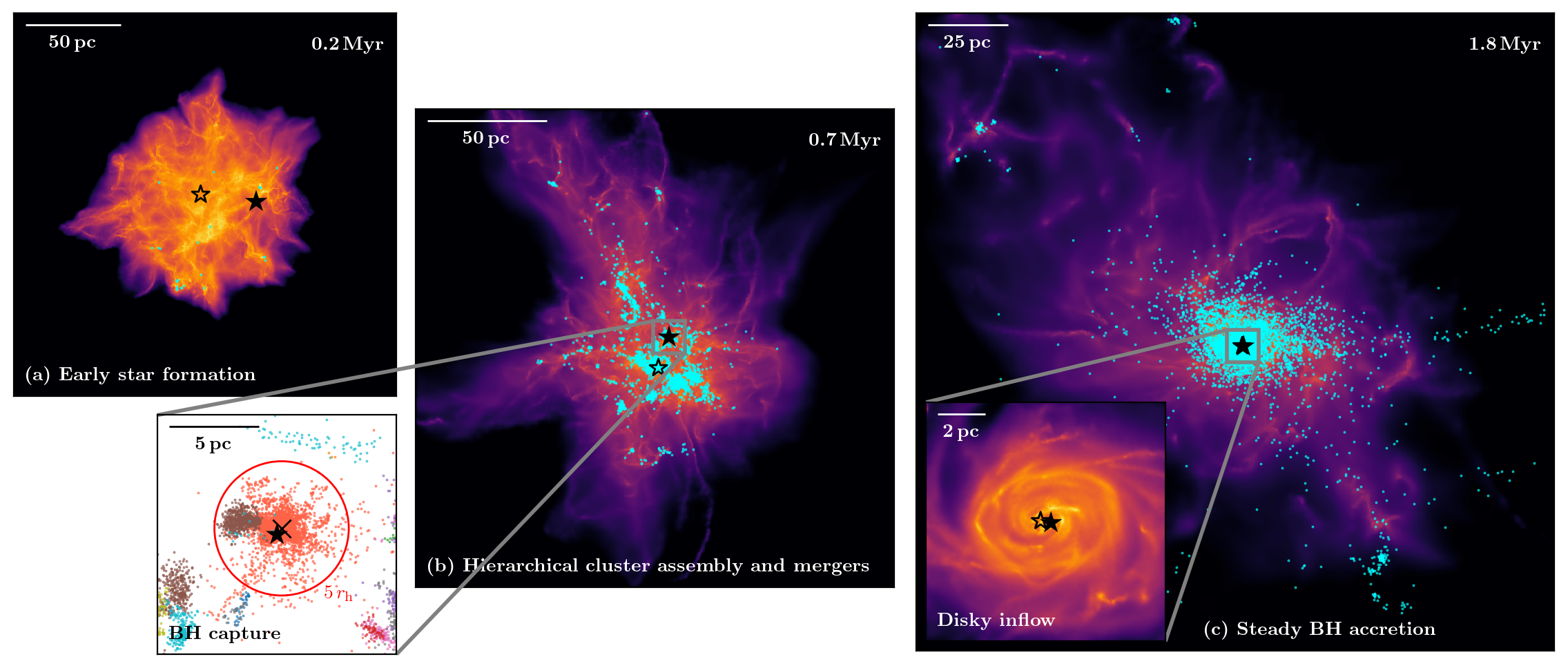}
    \vspace{-20 pt}
    \caption{Visualization of the simulation. We show the gas morphology (column density in logarithmic scales) and positions of star particles (colored dots; only a subset is displayed for aesthetic purposes) and two selected BHs (black stars) at different stages of the simulation. \textbf{Panel (a)}: Early star formation ($t \ll t_{\rm ff}$), where seed BHs wander in the GMC. \textbf{Panel (b)}: Hierarchical cluster assembly and mergers ($t \sim t_{\rm ff}$), where bursty star formation happens but the GMC is bound due to strong self-gravity. Massive sub-clusters form and one of the BHs is ``captured'' by (gravitationally bound to) a cluster (red dots in the inserted zoom-in panel, where the radius of the red circle equals 5 half-mass radii of the cluster). \textbf{Panel (c)}: Proto-bulge formation and steady BH accretion ($t\gtrsim 2\,t_{\rm ff}$), where subclusters merge into the final massive cluster, and the seed BH captured by the subcluster is migrated to the center. The inserted zoom-in panel shows the gas morphology around the BHs, which suggests convergent inflow and disk structure.
    %The inserted zoom-in panel shows the convergent inflow and disk structure near the BHs (adapted to the face-on direction), along with magnetic field lines with a strong toroidal component. Trajectories of the two BHs (cyan lines) suggest their migration to the center.
    %\pfh{Very nice figure. ICs panel doesn't add much, just need one ``early'' one ``late'' like other two panels. Since you talk so much about star sub-clusters, a map of the stars would be very helpful here. Maybe a density map at the same times and scales, so same image in both cases but just of stellar density, would really help see how it fragments then merges into one big cluster at the end.}
    }
    \label{fig:visualization}
\end{figure*}

\section{Simulation}
\label{sec:simulations}

The simulation is one of the suite in \citet{ShiKremerHopkins_2024arXiv240512164S}, where the numerics are described in detail, so we only briefly summarize the salient properties here. We utilize the magnetohydrodynamics (MHD) code \textsc{gizmo}\footnote{\url{http://www.tapir.caltech.edu/~phopkins/Site/GIZMO.html}} with MHD solved using the Meshless Finite Mass (MFM) method \citep{Hopkins_2015MNRAS.450...53H,HopkinsRaives_2016MNRAS.455...51H} and self-gravity using a tree-particle mesh method with self-consistent adaptive force softenings \citep{HopkinsNadlerGrudic_2023MNRAS.525.5951H}. In addition to self-gravity and (ideal) MHD, the code includes radiative cooling and heating, star formation, and stellar feedback, following the treatments in \citet{GrudicHopkinsFaucher-Giguere_2018MNRAS.475.3511G,ShiGrudicHopkins_2021MNRAS.505.2753S,ShiKremerGrudic_2023MNRAS.518.3606S,ShiKremerHopkins_2024arXiv240512164S}. In detail stars form from gas which is dense, molecular, rapidly-cooling and Jeans-unstable below the resolution limit and stars subsequently influence the medium via 5-band (FUV through IR) radiation (heating and photon momentum), stellar mass-loss, and supernovae, with rates tabulated assuming a well-sampled universal stellar initial mass function per the FIRE-2 implementation \citep{HopkinsWetzelKeres_2018MNRAS.477.1578H,HopkinsWetzelKeres_2018MNRAS.480..800H} of the Feedback In Realistic Environments (FIRE) physics \citep{HopkinsKeresOnorbe_2014MNRAS.445..581H}. Previous studies have shown this predicts reasonable properties of star formation, molecular clouds, and star clusters on galactic \citep{OrrHaywardHopkins_2018MNRAS.478.3653O,OrrHaywardMedling_2020MNRAS.496.1620O,OrrHatchfieldBattersby_2021ApJ...908L..31O,BenincasaLoebmanWetzel_2020MNRAS.497.3993B,KeatingRichingsMurray_2020MNRAS.499..837K} and individual star-cluster \citep{GrudicHopkinsFaucher-Giguere_2018MNRAS.475.3511G,GrudicGuszejnovHopkins_2018MNRAS.481..688G,GuszejnovGrudicOffner_2020MNRAS.492..488G,GrudicKruijssenFaucher-Giguere_2021MNRAS.506.3239G,GrudicHafenRodriguez_2023MNRAS.519.1366G,RodriguezHafenGrudic_2023MNRAS.521..124R,BruelRodriguezLamberts_2024A&A...686A.106B} scales.

%and follow the same treatment of star formation as previous works \citep{GrudicHopkinsFaucher-Giguere_2018MNRAS.475.3511G,ShiGrudic2021,ShiKremerGrudic_2023MNRAS.518.3606S,ShiKremerHopkins_2024arXiv240512164S}. Multiple physics are included in the simulation, including self-gravity, radiative cooling and heating, star formation, and feedback. Star formation and feedback are calculated with IMF-averages assemble, which follow the treatment of the FIRE-2 implementation of the Feedback In Realistic Environment (FIRE) framework \citep{HopkinsWetzel2018a}. The treatment is proven successful in recovering information about star formation and star cluster dynamics \citep{GrudicHopkinsFaucher-Giguere_2018MNRAS.475.3511G,GrudicGuszejnovHopkins_2018MNRAS.481..688G,GrudicKruijssen2021}.

In order to achieve a controlled experiment, we consider a non-cosmological idealized experiment of an initially turbulent, magnetized cloud complex, with some initial BH seed population. The complex has initial mass $10^8\,\rm M_\odot$ (all gas), spherical radius $50 \rm pc$, so surface density $\Sigma_{\rm gas}\sim 10^{4} \rm M_\odot yr^{-1}$, with initial turbulent virial parameter of unity (and initial turbulent/zero mean magnetic fields with magnetic energy equal to $1\%$ of gravitational energy), chosen to be representative of extreme but well-known conditions in starburst galaxy nuclei or high-redshift star-forming gas clumps in massive galaxies \citep{SwinbankSmailSobral_2012ApJ...760..130S,IzumiKawakatuKohno_2016ApJ...827...81I,ScovilleMurchikovaWalter_2017ApJ...836...66S,TacconiGenzelSaintonge_2018ApJ...853..179T}. We randomly distribute BH seeds spatially throughout the cloud, with initial masses uniformly distributed in $2 < \log_{10}({M}_{\rm seed}/{\rm M}_{\odot}) < 4$ and isotropic random velocities with the cloud virial velocity dispersion. Once the simulation begins, seeds can capture gas that is strictly bound to the BH (including thermal, magnetic, and kinetic energies) interior to the BH sink radius \citep{BateBonnellPrice_1995MNRAS.277..362B,ShiKremerGrudic_2023MNRAS.518.3606S}, which here is $\sim 0.1$\,pc (sufficient to easily resolve the Bondi radius of the relevant gas for accretion even at warm molecular temperatures, and the tidal radius of clumps/cores to the BHs, though of course much larger than the BH gravitational radii) comparable to state-of-the-art ``hyper-resolution'' simulations from ISM/galaxy scales \citep{Angles-AlcazarQuataertHopkins_2021ApJ...917...53A}. Since this still leaves the inner accretion disk unresolved, {we use a sub-grid model \citep[for detailed description see][]{ShiKremerHopkins_2024arXiv240512164S} to approximate the rate of this captured gas accreting onto the BH's inner disk ($\dot{M}_{\rm BH,0} \approx M_{\rm gas,\,disk}/t_{\rm dep}$}) with some disk depletion time following a \citet{ShakuraSunyaev_1973A&A....24..337S}-like scaling. {Apart from the mass accretion rate $\dot M_{\rm BH}$ that eventually contributes to the BH's mass growth, some mass $\dot{M}_{\rm w} = \eta_{\rm w} \dot{M}_{\rm BH} = \eta_{\rm w}/(1+\eta_{\rm w}) \dot{M}_{\rm BH,0}$} is assumed to be ejected in winds and bipolar jets following the methods in \citet{TorreyHopkinsFaucher-Giguere_2020MNRAS.497.5292T,SuHopkinsBryan_2021MNRAS.507..175S} with effective large-scale (emergent from un-resolved sales) velocity $v_{\rm j}$, while a fraction $\epsilon_{\rm r}$ emerges as radiation with the \citet{ShenHopkinsFaucher-Giguere_2020MNRAS.495.3252S} template spectrum. We set $\eta_{\rm w}=1$, $v_{\rm j}=400\,\rm km\,s^{-1}$, $\epsilon_{\rm r}=10^{-5}$. These are chosen to be representative of properties of super-Eddington ``slim disk'' models accreting at the hyper-Eddington rates we will find \citep{WataraiFukueTakeuchi_2000PASJ...52..133W,JiangStoneDavis_2019ApJ...880...67J}, but a parameter study of the effects of these choices is the subject of \citet{ShiKremerHopkins_2024arXiv240512164S}. We run the simulation for $1.8$\,Myr (a few times the initial global complex free-fall time), at which point we show the system is in quasi-steady state, with most star formation and BH growth complete.

To build a star sub-cluster merger history in post-processing, we first identify gravitationally bound star clusters in each snapshot.\footnote{\url{https://github.com/mikegrudic/Phinder}} To check the inheritance relationship between cluster $i$ in one snapshot and cluster $j$ in another one, we use a ``similarity measure'' which is defined as $s_{ij} = [n_{ij}/\min (n_i, n_j)]\cdot [2\sqrt{n_i n_j}/(n_i+n_j)]$, where $n_{i}$ ($n_j$) is the number of particles in the cluster $i$ ($j$), and $n_{ij}$ is the number of common particles between the two clusters. For a cluster in one snapshot, we iterate over clusters in the next snapshot to rank and determine the most possible successor (above a preset threshold). A ``merger tree'' is constructed after iterating over all clusters and all snapshots (densely sampling the local dynamical times) in the simulation.

\section{Proto-bulge formation and black hole capture and migration}
\label{sec:migration}

Fig.~\ref{fig:visualization} shows the gas and stellar morphology 
%(the column density in logarithmic scales) 
at different evolution stages in the simulation.
%: the initial condition (0\,Myr), the free-fall time ($t_{\rm ff}$, 0.6\,Myr), and the simulation time limit ($3t_{\rm ff}$, 1.8\,Myr). 
The simulation starts with a smooth density profile with initial turbulence seeded. Then gravitational collapse occurs and the medium becomes more turbulent, creating dense clumps, shocks, and filaments. 
Star formation begins in these dense regions well before one global free-fall time. The high surface density (or equivalently ``gravitational pressure'' or acceleration) scale is critical: it means the gas is not fully disrupted by ``early'' (pre-SNe) stellar feedback, despite turning most of its initial gas mass into stars, owing to the strong self-gravity. That in turn means that the bound-cluster star formation efficiency in the various sub-clumps is large (i.e.\ we form dense, bound subclusters, and not an unbound, low-density open cluster/association), consistent with both analytic expectations and results from a wide variety of simulations \citep{FallKrumholzMatzner_2010ApJ...710L.142F,GrudicHopkinsQuataert_2019MNRAS.483.5548G,GrudicBoylan-KolchinFaucher-Giguere_2020MNRAS.496L.127G,WadaPapadopoulosSpaans_2009ApJ...702...63W,HopkinsTorreyFaucher-Giguere_2016MNRAS.458..816H,HopkinsWellonsAngles-Alcazar_2022MNRAS.510..630H}.

%Star formation is also initialized significantly before the first free-fall time. At later times, as the right panel shows, the GMC is not disrupted by stellar feedback due to the cloud's strong self-gravity \citep{GrudicHopkinsFaucher-Giguere_2018MNRAS.475.3511G}. Instead, there is a convergent inflow toward the center of the cloud. Further zoom-in reveals the inflow builds up into disk structure with a size of $\sim 10\,\rm pc$. On top of the gas density, we present the magnetic field lines in the zoom-in panel, which have strong toroidal components.As black stars, we show the two BH seeds that experience the most significant accretion. We find that both BHs migrate to the center where the gas converges. More interestingly, the trajectory of the BH has a sharp turn before it spirals to the center. We expand more details of the BH migration and the disk structure in the next two sections.

\begin{figure*}
    \centering
    % \includegraphics[width=1.05\linewidth]{plots/cluster-mergers-trajectory.pdf}
    % \vspace{-15 pt}
    % \includegraphics[width=1.05\linewidth]{plots/cluster-mergers-mass.pdf}
    % \vspace{-15 pt}
    \includegraphics[width=\linewidth]{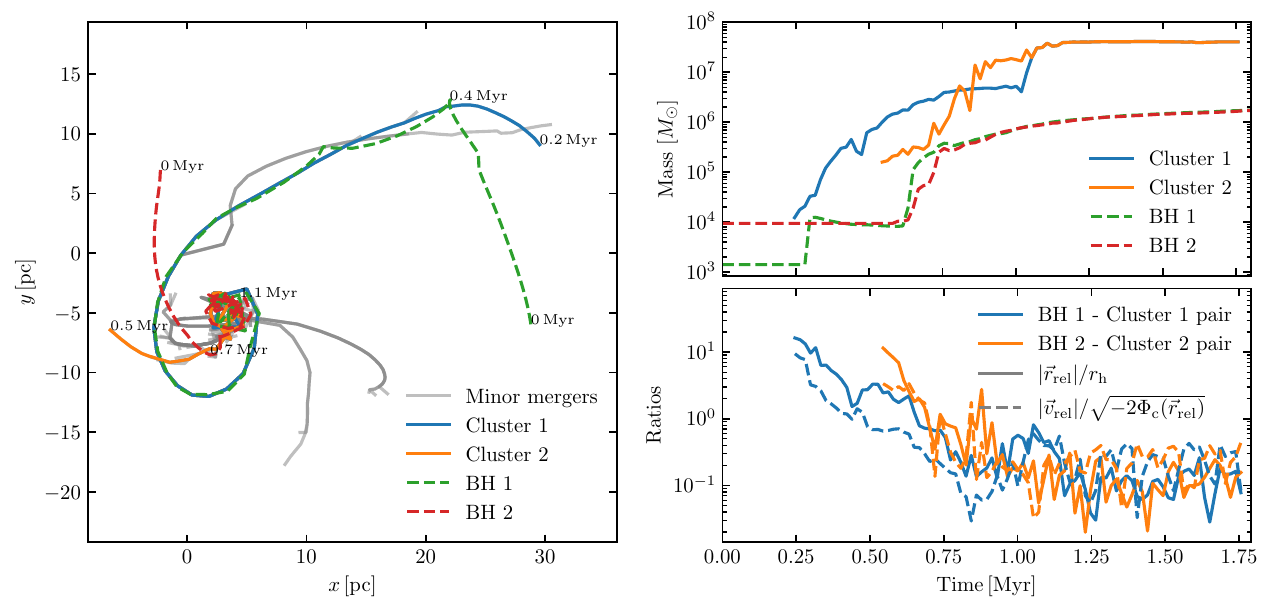}
    \vspace{-20 pt}
    \caption{BH capture, migration, and growth as subclusters merge. 
    \textbf{Left}: Trajectories of the centers-of-mass of the two subclusters (1,2; {\it solid}) which capture the two BHs (1,2; {\it dashed}) from Fig.~\ref{fig:visualization}. The BHs are ``captured'' (become bound to the subcluster) at the times labeled, well before they reach the center, and are carried to the center by the merging subclusters. Minor subclusters merged into these two massive clusters are tracked with gray lines.
    \textbf{Upper right}: Mass evolution of the subclusters and BHs. Clusters acquire mass initially through star formation then hierarchical merging, which completes (and the system relaxes) by $\sim 1\,$Myr. 
    The BHs grow in early ``jumps'' as they accrete clumps near their subcluster centers, then at a steady rate $\sim 1-5\rm M_\odot yr^{-1}$ after $\sim 0.7$\,Myr from the nuclear disk.
    \textbf{Bottom right}: Boundedness/sinking of each BH in its sub-cluster. We show the spatial ($|\mathbf{r}_{\rm rel}|$; {\it solid}) and velocity ($|\mathbf{v}_{\rm rel}|$; {\it dashed}) separation between BH and its subcluster center-of-mass, normalized by the subcluster half-mass radius $r_{\rm h}$ or escape velocity $\sqrt{-2\Phi_{\rm c}(\mathbf{r}_{\rm rel})}$. The BHs sink and become more confined as their subclusters grow and merge.}
    \label{fig:mergers}
\end{figure*}

%\pfh{Re: Fig 2: The name ``cluster merger track'' is awkward. I would change the label to something simpler like ``subcluster 1''. and for the BHs change label to match lower panels to be just ``BH 1'' and ``BH 2''. What is the grey line in the top panel of the Figure?}

Given the high star formation efficiency, we see these form into hierarchical sub-clusters that merge rapidly, as in local young massive clusters \citep{GrudicGuszejnovHopkins_2018MNRAS.481..688G,LiVogelsbergerMarinacci_2019MNRAS.487..364L,GuszejnovMarkeyOffner_2022MNRAS.515..167G}. 
%Clusters form hierarchically in GMCs \citep{GuszejnovMarkey2022}. Moreover, for the dense GMC we simulated, the gravitational force is dominant over feedback, which means that gravitational interactions between stars can be more significant. 
In Fig.~\ref{fig:mergers} we show the trajectories of (sub)cluster centers,  following their merger history. We can see that clusters merge rapidly after their formation and migrate to the center. We also plot trajectories of the two BH seeds with the most significant mass accretion (same as those in Fig.~\ref{fig:visualization}), and highlight cluster merger tracks that interact with them. BH 1 encounters cluster merger track 1 at $\sim 0.4\,\rm Myr$ and becomes captured by that cluster; it then oscillates along the cluster's trajectory before settling down in the cluster. The BH-cluster pair then migrate to the center at $\sim 1.1 \,\rm Myr$. A similar process occurs for the second BH-cluster pair, though this capture happens at $\sim 0.7\,\rm Myr$ and is much closer to the center. These two clusters later merge at $\sim 1.1\,\rm Myr$, rapidly bringing the BHs together to form a binary. We do not allow the BHs to merge since we do not resolve the radii ($\ll 0.1\,$pc) where gravitational wave emission would become important to coalescence.
%\kyle{Just curious, do we allow the black holes to merge? Or has sufficient time elapsed to enable a BH merger? Not sure if we need to discuss this or not...}

The merger history of clusters is also reflected in their mass evolution, as shown in the top right panel of Fig.~\ref{fig:mergers}. The two clusters merge into a very massive cluster ($\sim 4\times 10^7\,\rm M_\odot$) at $\sim 1\,\rm Myr$, after which the clusters' mass growth stalls. We also plot the mass evolution of the two BHs which grow most effectively. Both BHs reach $\sim 10^4\,\rm M_\odot$ by $\sim 0.5\,\rm Myr$, with most of this ``early'' accretion occurring when a dense clump become gravitationally captured by the BH, similar to the behavior described in \citet{ShiKremerGrudic_2023MNRAS.518.3606S}. Note that this early, ``stochastic'' growth is shown in \citet{ShiKremerGrudic_2023MNRAS.518.3606S} to be essentialy independent of seed mass over the range $M_{\rm seed} \sim 100-10^{4} \rm M_\odot$, consistent with what we see here (where one seed has about the median mass of our random seeding, while the other happens to be towards the high-mass end of our initial seed spectrum). What is more important is that these seeds (of the $\sim 100$ initial seeds placed randomly throughout the complex) happened to be at the ``right place at the right time'' to be captured and reach the center of their sub-cluster quicky. After $\sim 0.75\,\rm Myr$, the two BHs have reached the cluster center and the cluster has started to coalesce and relax, so the large-scale accretion flow (discussed below) becomes more stable and the BHs continue to grow almost linearly in time, i.e. at a constant $\dot{M}_{\rm BH}\sim 2-5\,\rm M_\odot/yr$. They finally both reach $\sim 2\times 10^6\,\rm M_\odot$ at the end of the simulation. Note that at all times the BHs are much less massive than their host (sub)cluster, so their large-scale sinking and coalescence to the center, over a distance of $\sim 50\,$pc, is enabled by the rapid dynamical hierarchical subcluster merging. If the BH seeds instead had to sink to the center of the final, relaxed/smooth star cluster as isolated objects (i.e.\ via classical dynamical friction), the dynamical friction time would be $>10^{10}$\,yr. Instead, since the subclusters hierarchically merge in $\mathcal{O}(1)$ mass ratios, the BHs can be carried to the center in of order one dynamical time \citep[][and references therein]{GuszejnovMarkeyOffner_2022MNRAS.515..167G}.

Through evolution, each cluster also changes its mass $M_{\rm c}$ and half-mass radius $r_{\rm h}$. For the two clusters highlighted in Fig.~\ref{fig:mergers}, mass growth, and accompanying deepening of their potential well, enhance capture of their respective BHs. This is illustrated in the bottom right panel of Fig.~\ref{fig:mergers}, where we track the relative distance ($\mathbf{r}_{\rm rel}\equiv \mathbf{r}_{\rm BH} - \mathbf{r}_{\rm cluster}$, normalized to $r_{\rm h}$, where $\mathbf{r}_{\rm cluster}$ is the cluster's gravitational center) and velocity ($\mathbf{v}_{\rm rel} = \dd \mathbf{r}_{\rm rel}/\dd t$, normalized to the escape velocity at the BH position $v_{\rm esc}=\sqrt{-2\Phi_{\rm c}(\mathbf{r}_{\rm rel})}$, {where we calculate the gravitational potential $\Phi_{\rm c} (\mathbf{r}_{\rm rel}) \equiv - \sum_{i} G m_i/|\mathbf{r}_{\rm BH} - \mathbf{r}_i|$ at the BH position from all stars in the cluster at that time}). 
%Since clusters evolve with time, we rescale each quantity with characteristic units. For distance, we choose the half-mass radius $r_{\rm h}$, which characterizes the size of the cluster. For velocity, we use the escape velocity of the potential well at the BH's position, defined as 
%$v_{\rm esc}=\sqrt{-2\Phi_{\rm c}(\mathbf{r}_{\rm rel})}$, where $\Phi_{\rm c}(\mathbf{r}) = -\sum_i G m_i/|\mathbf{r}-\mathbf{r}_{i}|$ is the gravitational potential due to all particles in the star cluster. 
We see distance and velocity decay in $\lesssim 1\, \rm Myr$ (the same is true if we measure the specific angular momentum), for both the BHs within their sub-clusters and the subclusters with respect to the parent cloud/cluster.
%We find that each BH sinks to the companion cluster's center in $\sim 1\,\rm Myr$. This trend is also true for the specific angular momentum ($\mathbf{j}_{\rm rel}\equiv \mathbf{r}_{\rm rel} \times \mathbf{v}_{\rm vel}$) if rescaled with $\sqrt{G M_{\rm c} r_{\rm h}}$. 
But we see that capture is not just due to a chance encounter between a BH and subcluster at low speeds (which would be rare). Rather Fig.~\ref{fig:mergers} shows that the two BH-cluster pairs are not gravitationally bound initially, and indeed the relative velocities of the BHs to clusters actually increase (from $\sim 50$ to $\sim 200\,\rm km\,s^{-1}$) from early to late times, but the ratios of kinetic to potential energy become $<1$ and the BHs are captured at $0.4$ and $0.7$\,Myr, the same times that their clusters grow rapidly in mass, increasing $|\Phi_{\rm c}|$ and the subcluster escape velocity rapidly relative to the (relatively weaky-evolving) $v_{\rm rel}$.
%We {emphasize} that the capture is not only due to a {chance} encounter between a BH and its companion cluster but is largely amplified by the deepening of the cluster's potential well. As shown in the bottom panel of Fig.~\ref{fig:mergers}, the two BH-cluster pairs are not gravitationally bound at the beginning, but the ratio of kinetic energy over potential energy is lower than 1 at $\sim 0.4\,\rm Myr$ and $\sim 0.7\,\rm Myr$ respectively, which is approximately the time when the BHs are captured by their companion cluster (see the upper panel of Fig.~\ref{fig:mergers}). However, the actual relative velocity between the BH and the cluster grows from $\sim 50\,\rm km/s$ to $\sim 200\,\rm km/s$ at the late time, meaning that the mass assembly history of clusters is critical.

\begin{figure*}
    \centering
    \includegraphics[width=.8\linewidth]{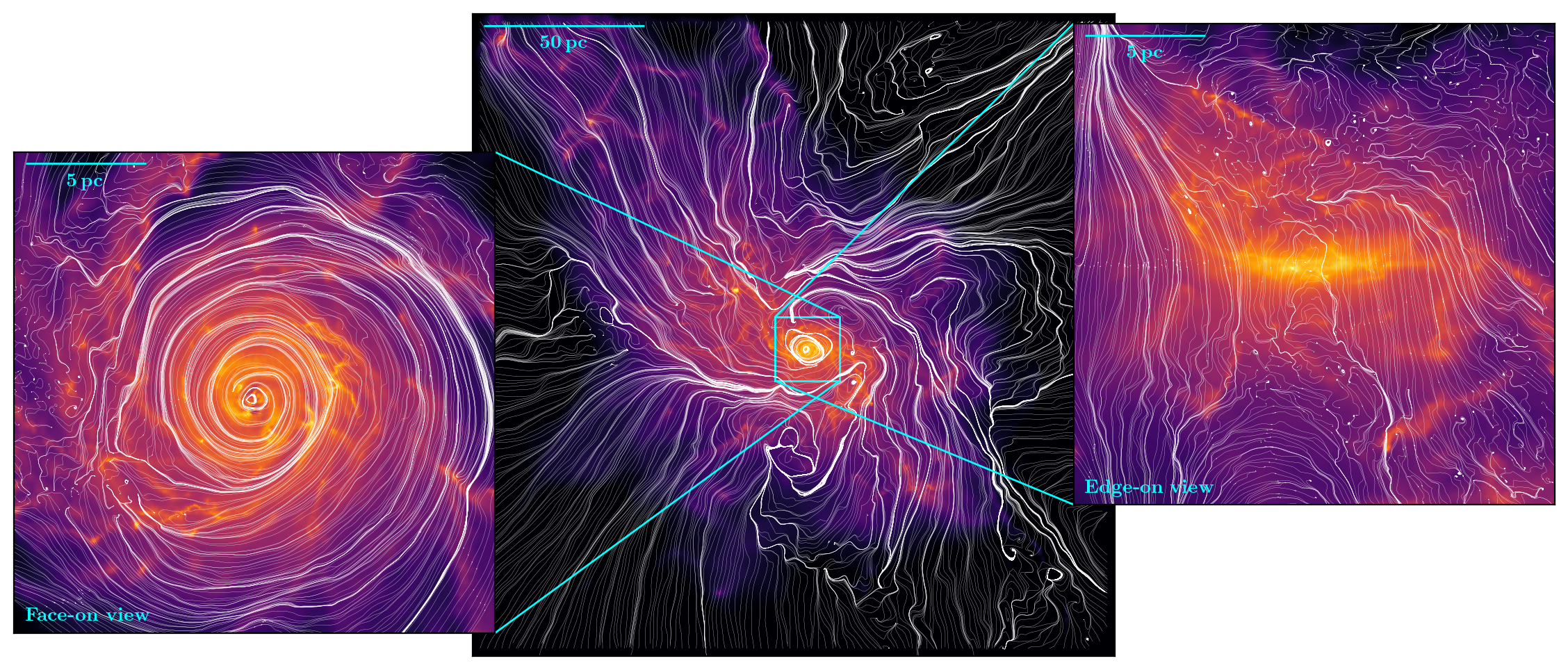}
    \vspace{-10 pt}
    \caption{Magnetic field lines (near the midplane of each box) and gas morphology (logarithm of the column density) at the late stage of the simulation. The main panel suggests the large-scale magnetic field is advected to smaller scales due to gravitational collapse throughout the simulation. The inserted zoom-in panels focus on the disk structure, adapted to face-on (left panel) and edge-on (right panel) views. The face-on view shows a dominant toroidal magnetic field spiraling inward anti-clockwise, while the edge-on view shows the (weaker) poloidal magnetic field lines threading through the disk structure.
    }
    \label{fig:magnetic_field}
\end{figure*}

The final merged star cluster has a mass of $\sim 5\times10^{7} \rm M_\odot$, but with a dense central mass concentration (half-mass radius of $\sim 3$\,pc) and isothermal-sphere like density profile $\rho_{\ast} \propto r^{-2}$ steepening to $r^{-4}$ at larger radii, after undergoing violent relaxation from multiple hierarchical subcluster mergers. This means the characteristic circular velocity is large, $\sim 200\,\rm km\,s^{-1}$ (see \S\ref{sec:disk}), and the cluster lies in-between massive typical nearby star clusters and galactic bulges in most structural and kinematic scaling relations (where the two are known to form a continuous family; see \citealt{HopkinsMurrayQuataert_2010MNRAS.401L..19H,GrudicHopkinsQuataert_2019MNRAS.483.5548G}). Indeed, these are very similar stellar masses, sizes, surface densities, velocity dispersions, and density profiles to known massive nuclear star clusters and ultra-compact dwarfs observed \citep{GehaGuhathakurtavanderMarel_2002AJ....124.3073G,WalchervanderMarelMcLaughlin_2005ApJ...618..237W,HaseganJordanCote_2005ApJ...627..203H,EvstigneevaGreggDrinkwater_2007AJ....133.1722E}. So given the initial gas and final stellar properties, this suggests a picture where these ``proto-bulges'' form from sufficiently high-density gas complexes in high-redshift massive galaxy progenitors, which rapidly fragment into sub-clusters that can capture (even relatively modest mass) BH ``seeds'' and hierarchically merge in a dynamical time (as argued from some cosmological simulations in e.g.\ \citealt{MaHopkinsBoylan-Kolchin_2018MNRAS.477..219M,MaGrudicQuataert_2020MNRAS.493.4315M}), forming a deep, centrally-concentrated potential which traps and brings together the BHs.

It is worth mentioning that hierarchical star cluster assembly is a process with ample dynamics that may produce massive seed BHs. For example, \citet{RantalaNaabLahen_2024MNRAS.531.3770R} found that the process may boost the formation of very massive stars (VMS) through repeated star mergers, which can be progenitors of BH seeds. Still, for gas-rich environments which this work is focused on, BH accretion can play an important role in the ``proto-bulge'' system. We expand on these details in the next section.

% \begin{figure}
%     \centering
%     \includegraphics[width=\linewidth]{plots/BH-cluster-pairs.pdf}
%     \vspace{-20 pt}
%     \caption{Sinking of BHs in their evolving parent cluster. From top to bottom, we show the relative distance, velocity, and specific angular momentum between the BH and the cluster, all rescaled with cluster mass profiles which are also evolving. Despite the absolute value of kinetic energy and angular momentum of each BH do not change significantly, BHs become bound and sink to clusters through the evolution of clusters.
%     }
%     \label{fig:BH-cluster}
% \end{figure}

\section{Sustained Hyper-Eddington Accretion Through a Hyper-Magnetized Disk}
\label{sec:disk}

\begin{figure*}
    \centering
    \includegraphics[width=\linewidth]{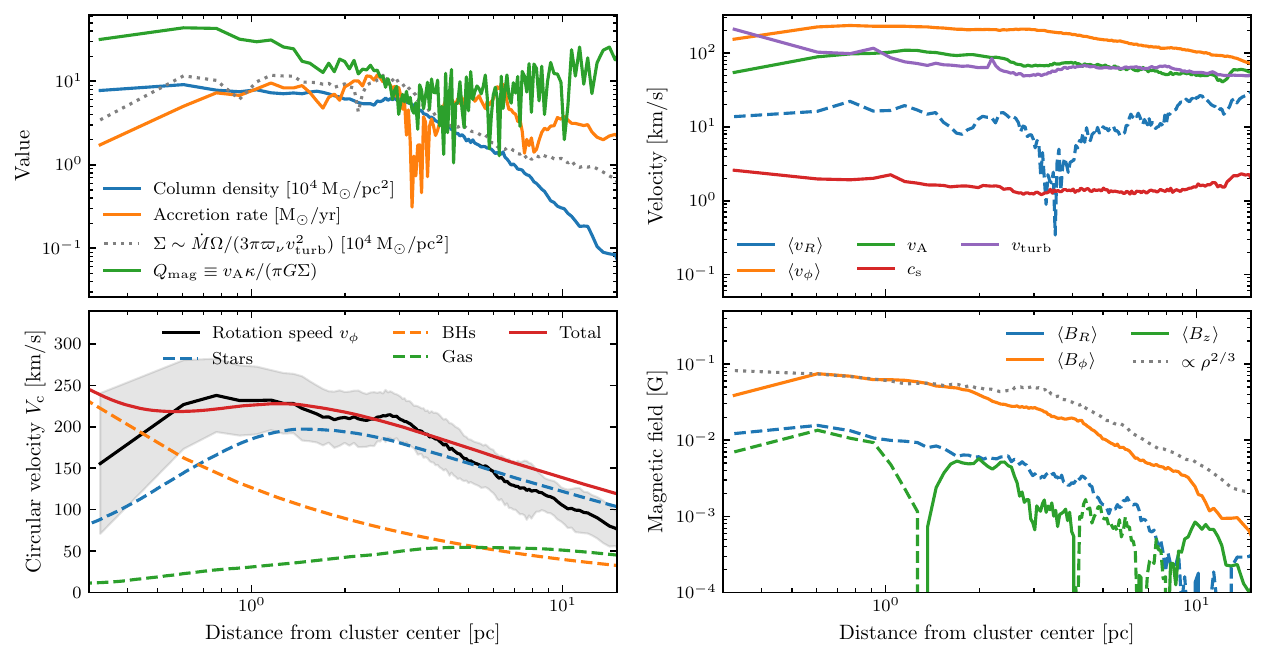}
    \vspace{-20 pt}
    \caption{
    Dynamics and magnetic field of the disk. 
    \textbf{Upper left}: surface/column density, and mass accretion rate, and magnetic Toomre $Q_{\rm mag}$ profiles of the disk (measured in cylindrical annuli). 
    We also plot the predicted surface density $\Sigma(R)$ for the analytic hyper-magnetized disk models in \citet{HopkinsSquireQuataert_2024OJAp....7E..20H}, assuming an accretion rate $\dot{M}/\varpi_{\nu}=30\,{\rm M_{\odot}/yr}$.
    \textbf{Lower left}: mean rotation velocity $v_\phi$ of the gas measured from the simulation ({with the gray band representing the standard deviation}), and the total circular velocity $\equiv \sqrt{G\,M_{\rm tot}(<R)/R}$ with the contribution $\equiv \sqrt{G\,M_{i}(<R)/R}$ the mass of stars, BHs, and gas. 
    \textbf{Upper right}: average radial and azimuthal velocities (\textit{dashed} if negative) of the gas in the disk, and Alfven speed ($v_{\rm A}$), thermal sound speed ($c_{\rm s}$), and root-mean-square turbulent velocities ($v_{\rm turb}$). 
    \textbf{Lower right}: strength of the mean radial, toroidal, and poloidal magnetic field (\textit{dashed} if negative). The overall magnetic field strength is in agreement with the flux-freezing assumption ($|\mathbf{B}|\propto \rho^{2/3}$, \textit{dotted}).
    %Dynamics and magnetic field of the disk. \textbf{Panel 1} (from top to bottom): surface density and mid-plane mean density of the profile of the disk. \textbf{Panel 2}: circular velocity $v_\phi$ of the disk and the Keplerian velocity fitting from stars, BHs, and gas. \textbf{Panel 3}: the radial, circular, and perpendicular velocity components of the gas inflow in the disk, and characteristic speeds for different physics including magnetic field ($v_{\rm A}$), thermal pressure ($c_{\rm s}$), and turbulence ($\sigma_z$). \textbf{Panel 4}: strength of the radial, toroidal, and poloidal magnetic field (dashed lines if negative). \textbf{Panel 5}: plasma beta ($\beta$), electron fraction ($f_e$, including its scattering), and the ratio between viscous heating ($Q_{+}^{\rm vis}$) and blackbody cooling rates ($Q_{-}^{\rm BB}$).
    }
    \label{fig:disk}
\end{figure*}

As the clusters begin to dynamically ``settle'' after $\sim 0.4-0.7$\,Myr, we see an extended accretion disk appear and grow ``inside-out,'' beginning just outside the sink radius until ultimately extending to $\gtrsim 10$\,pc scales. 
Fig.~\ref{fig:visualization} plainly shows the disk in the last inserted panel, while Fig.~\ref{fig:magnetic_field} shows the magnetic field structures. We overplot the magnetic field vectors, which are clearly dominated by a dominant mean toroidal/azimuthal field, though there are also poloidal fields vertically threading through the disk. This results in the nearly-constant growth shown in Fig.~\ref{fig:mergers} at $\sim 1-10\,{\rm M}_{\odot}\,{\rm yr^{-1}}$ for the remaining $\gtrsim 1\,$Myr. From this and the BH masses, we see this is well above the Eddington limit which ranges from $\sim 0.006-0.04\,{\rm M}_{\odot}\,{\rm yr^{-1}}$ over the same time. We also see the disk extends well outside the BH radius of influence $R_{\rm ROI} = G M_{\rm BH}/\sigma_{\ast}^{2} \sim 0.1\,{\rm pc} (M_{\rm BH}/10^{6} {\rm M}_{\odot})\,(\sigma_{\ast}/200\,{\rm km\,s^{-1}})^{-2}$.

%At late times, our two BHs of interest grow linearly in mass due to the convergent inflow of gas arriving at the center of the cluster and settling into an accretion disk (see Fig.~\ref{fig:visualization}). In this section, we study the dynamics, magnetic field, and thermodynamics of the disk.

We summarize the gas disk properties in Fig.~\ref{fig:disk}. We confirm both that it is indeed a disk (primarily rotation supported with $\langle v_{\phi}\rangle \approx V_{\rm c}^{\rm total}=\sqrt{G M_{\rm tot}(<R)/R}$ out to $\gtrsim 10\,$pc), and that it clearly appears to be an example of the ``hyper-magnetized'' or ``flux-frozen'' disks proposed in \citet{HopkinsGrudicSu_2024OJAp....7E..18H,HopkinsSquireSu_2024OJAp....7E..19H,HopkinsSquireQuataert_2024OJAp....7E..20H}. The defining properties of these disks, all of which we see in Fig.~\ref{fig:disk}, are that magnetic pressure dominates in the midplane (plasma $\beta  \sim P_{\rm thermal}/P_{\rm B} \sim c_{s}^{2} / v_{A}^{2} \sim 10^{-3}$ here),\footnote{We have verified, and it is expected from simple dimensional arguments, that radiation pressure is negligible in the midplane at these (large) disk radii.} coming primarily from a mean toroidal field (with $\langle B_{\phi} \rangle$ much larger than either the mean radial/poloidal fields $B_{R}$, $B_{z}$ or turbulent fields), with trans (or weakly sub) Alfvenic turbulence ($v_{\rm turb} \sim \,v_{A}$ here).
%($v_{\rm turb} \sim \sigma_{z}^{\rm gas} \sim 0.3-1\,v_{A}$ here).
One can verify from Fig.~\ref{fig:disk} that the accretion rates through the disk are approximately constant ($\sim 1-10\,\rm M_\odot yr^{-1}$) with $R$, with values consistent with the measured mean radial $\langle v_{R} \rangle$, and Maxwell+Reynolds stresses directly extracted from the simulation (note the expected $\langle B_{R} \rangle-\langle B_{\phi} \rangle$ anti-correlation, discussed in \citealt{HopkinsSquireSu_2024OJAp....7E..19H}, which ensures a strong Maxwell stress of the desired sign) as well as that the (thick) vertical height behaves as expected from magnetic+turbulent support ($H/R \sim \sigma_{z}/V_{\rm c} \sim v_{A}/V_{\rm c}\sim 0.3-1$), and the midplane density scales accordingly ($\rho \sim \Sigma_{\rm gas}/2H$). For a detailed analysis of these properties in such disks, we refer to \citet{HopkinsSquireSu_2024OJAp....7E..19H}. 

The magnetic field strengths appear to be consistent with growth via flux-freezing/advection of magnetic flux from the diffuse gas into and throughout the disk: $B_{\phi}$ rises approximately as $r^{-1}$ from a couple hundred $\mu{\rm G}$ (the value of the turbulent field in the initial conditions, comparable to that observed in gas with a mean density similar to the initial cloud value $n \sim 10^{4}\,{\rm cm^{-3}}$ in the ISM; see \citealt{Crutcher_2012ARA&A..50...29C,PonnadaPanopoulouButsky_2022MNRAS.516.4417P}) to $\sim 0.1$\,G on sub-pc scales. If we follow the component-wise analysis in \citet{HopkinsSquireSu_2024OJAp....7E..19H}, this amplification as well as the $B_{\phi}-B_{R}$ anticorrelation are consistent with the mean-field Lagrangian induction equation, as the gas is captured and compressed (and the fields ``wound up'' toroidally) as it accretes.\footnote{We can also verify that our assumption of ideal MHD is valid. The microphysical Spitzer/Braginskii conduction/viscosity coefficients are tiny under these gas conditions (equivalently particle mean free paths are extremely small compared to $\sim H$). And comparing non-ideal Ohmic/Hall/ambipolar resistivities \citep[e.g.][]{KunzMouschovias_2009ApJ...693.1895K} to the turbulent diffusivities $\sim v_{t}\,H$ shows that the microphysical non-ideal terms are smaller by at least $\sim 5-10$ orders of magnitude, owing to a combination of the relatively low gas densities and high ionization fractions $f_{\rm ion}\sim 0.01$ given the extreme stellar irradiation of the low-density disk.} As well, we overplot the specific analytic model dependence of $\rho(R)$, $B(R)$ from \citet{HopkinsSquireQuataert_2024OJAp....7E..20H}, after modifying their model to account for the non-Keplerian potential seen here (below), and see reasonable agreement.

As in \citet{HopkinsGrudicSu_2024OJAp....7E..18H}, the strong magnetic field appears to stabilize the disk against fragmentation and star formation. We do not see fragmentation and the star formation rate after the disk forms is small (Fig.~\ref{fig:mergers}) compared to BH growth rates. Indeed, from Fig.~\ref{fig:disk} we can see that the Toomre $Q$ parameter accounting for magnetic support ($Q_{\rm mag}\equiv v_{\rm A}\kappa/\pi G\Sigma$) is $\gg 1$ at all radii here (typically $\sim 20-50$), while if we included only thermal pressure ($Q_{\rm therm}\equiv c_{\rm s} \kappa/\pi G \Sigma$) we would obtain $Q_{\rm therm} \sim 0.1 \ll 1$ at all radii. 

The thermal properties themselves, while unimportant for the disk structure since $\beta \ll 1$ (and because we are well outside the near-horizon radii where most of the accretion disk emission originates), are reasonable. At these large ($\sim$\,pc) radii the disk self-heating flux $\sim (3/8\pi) \dot{M} \Omega^{2}$ is small compared to the flux from stellar irradiation. Assuming a cluster mass profile dominated by a young ($\lesssim 10$\,Myr) stellar population, with a well-sampled IMF, and standard dust composition as assumed in-code, the equilibrium dust temperature (equating the stellar flux absorbed by grains to their emission) is $\sim 400\,{\rm K}\,(V_{\rm c}/200\,{\rm km\,s^{-1}})^{2/5}\,(R/{\rm pc})^{-1/5}$, which at the densities here should be in rough equilibrium with the gas temperature. Indeed this appears to reasonably predict the thermal properties in Fig.~\ref{fig:disk}, though we note the disk is somewhat multi-phase with a mix of warm and cool atomic, cold molecular gas, and very compact HII regions around some stars.

There are some notable differences from the simulation presented in \citet{HopkinsGrudicSu_2024OJAp....7E..18H}. First,  the disk here forms under very different circumstances and in a very different regime of parameter space (in terms of BH mass, gas/galaxy properties, accretion rate, initial absolute magnetic field strength, etc.), and numerically the simulations here adopt a different MHD solver, treatments of star formation and stellar feedback (FIRE versus STARFORGE; \citealt{GrudicGuszejnovHopkins_2021MNRAS.506.2199G}), resolution, BH feedback treatment, and initial conditions. All of which argues that such disks should not be uncommon in high-accretion rate BHs. Second, and perhaps most notably, the disk in those papers formed entirely within the ROI, where we see it extend well past the ROI here at all times. Fig.~\ref{fig:disk} explicitly shows that stars dominate the potential outside $\gtrsim 0.5\,$pc even at the final time (when the BHs are most massive), with a somewhat flat, then again weakly declining $V_{\rm c}(R)$. This implies that there is nothing uniquely special about a Keplerian potential required to obtain such a disk, so long as there is a well-defined centrally-peaked potential, and the magnetic fields, gravitational stability, and other properties of the gas permit. Indeed we can still apply the analytic models for the predicted structure of such disks derived in \citet{HopkinsSquireQuataert_2024OJAp....7E..20H}, if we replace $\Omega$ in their scalings (which they took to be Keplerian) with the appropriate value here $\Omega = V_{\rm c}/r$. If we take their default model and make this replacement, we predict $-\partial \ln{B_{\phi}}/\partial \ln R \sim 8/9-4/3$ and  $-\partial \ln{\rho}/\partial \ln R \sim 4/3-2$ (with the shallower slopes around $\sim 0.5-3\,$pc, where $V_{c}\sim R^{0}$ deviates most from Keplerian), consistent with the behavior in Fig.~\ref{fig:disk}. This also has implications for the disk heating/thermal structure (above). Third, it is worth noting that this is formally a circum-binary disk, given the two nearly equal-mass BHs (Fig.~\ref{fig:mergers}). 

It is striking that the two BHs, once brought together at $\gtrsim0.7$\,Myr, rapidly converge to nearly identical mass growth rates in the interior of the circumbinary disk. This effect has been seen before in circumbinary disk simulations \citep{LaiMunoz_2023ARA&A..61..517L} and may explain populations of ``twin stars'' \citep{El-BadryRixTian_2019MNRAS.489.5822E}, but will clearly have important implications for BH mergers.

\section{Conclusions}
\label{sec:conlcusions}

\begin{figure*}
    \centering
    \includegraphics[width=\linewidth]{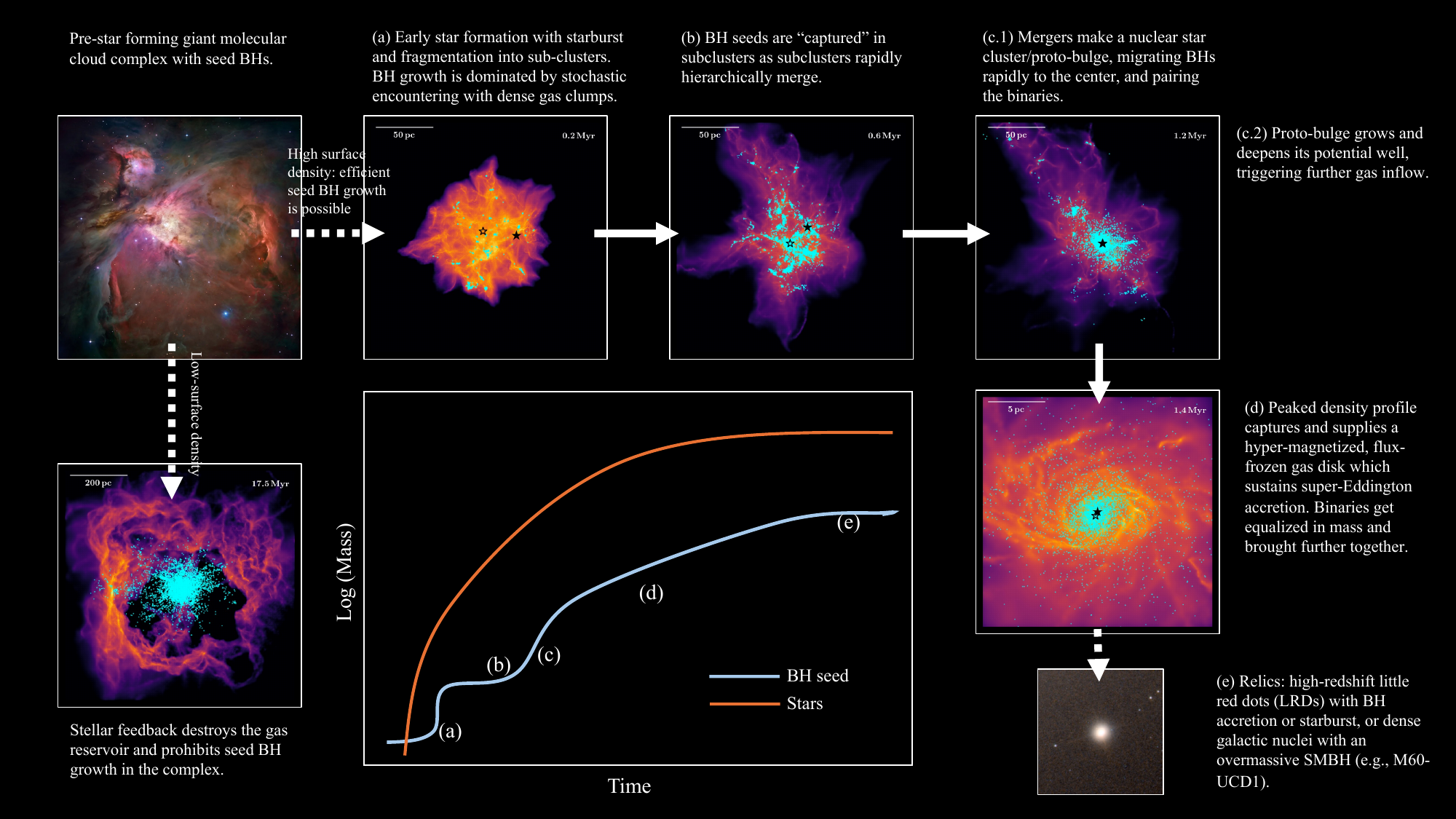}
    \vspace{-15pt}
    \caption{Illustration of this scenario of seed BH capture, growth, migration, and pairing in the star-forming GMC complex. The evolution of the star-forming complex is largely determined by the surface density, while low-surface-density clouds are expelled by stellar feedback, halting possible seed BH accretion (upper left $\to$ lower left). In high-surface-density clouds, seed BHs may undergo steps (a) $\to$ (e), corresponding to the qualitative trend of star formation and BH mass growth (lower middle panel; also see Fig.~\ref{fig:mergers}). \textit{Image credits} -- M42 (\textit{upper left panel}): NASA, ESA, M. Robberto (Space Telescope Science Institute/ESA) and the Hubble Space Telescope Orion Treasury Project Team; M60-UCD1 (\textit{panel e}): NASA, ESA, and the Hubble Heritage (STScI/AURA).% LRDs (panel e): adapted from Fig.~6 of \citet{kocevski:2024.little.red.dots.jwst}.
    }
    \label{fig:illustration}
\end{figure*}

We present a scenario for extremely rapid ($\sim 1\,$Myr) transition from seed-to-supermassive BHs in a forming proto-bulge or massive nuclear star cluster, with efficient growth and pairing of BH binaries. This unifies many theoretical ideas which have been proposed for different aspects of the problem, demonstrating for the first time all of them operating in concert in an MHD-thermochemical-star formation simulation. Our key conclusions, which define the steps in this scenario (also see Fig.~\ref{fig:illustration}), unfold as follows:
\begin{enumerate}

\item In a massive gas cloud complex of $\sim 10^{8}\rm M_\odot$ and surface density $\Sigma \sim 10^{4} \rm M_\odot pc^{-2}$, typical of starburst nuclei or ``clumps'' in massive high-redshift galaxies, fragmentation and star formation are very efficient and stellar feedback is weak, so many bound sub-clusters rapidly form. A small fraction of low-mass BH seeds can grow essentially ``stochastically'' to IMBH masses by encountering {dense}, small clumps of gas.

\item Those IMBHs are efficiently ``captured'' by their adjacent subclusters during this process. As the subclusters grow and merge the BHs become even more effectively confined.

\item Subclusters merge hierarchically in essentially one dynamical time of the complex ($<1\,$Myr). This builds a proto-bulge, with properties similar to observed massive, dense nuclear star clusters and UCDs, with a sharply centrally-concentrated (isothermal) stellar mass profile.

\item Carried by their subclusters, the trapped BHs migrate to the center in the same time, orders of magnitude faster than their dynamical friction time if they were isolated. This also produces efficient ``pairing'' down to $<1\,$pc scales of multiple BHs.

\item The proto-bulge provides a deep, centrally-concentrated potential which allows gas to be globally captured and retained in the system, with a coherent center and angular momentum, so it forms a large-scale accretion disk extending well beyond the BH radius of influence (to $\gtrsim$\,pc scales). Accretion transitions from stochastic/turbulent clump-clump encounters to disk feeding. 

\item As gas is captured with modest pre-existing turbulent magnetic fields, the toroidal field lines are stretched and form a magnetically-dominated, flux-frozen disk. The disk pressure comes from the toroidal field, with plasma $\beta \sim 10^{-3}$ in the midplane. This stabilizes the disk against catastrophic fragmentation and star formation and allows it to exist at all, while simultaneously ensuring strong Maxwell and Reynolds stresses that support extremely large accretion rates $\sim 1-10\,\rm M_\odot yr^{-1}$ for $\gtrsim\,$Myr timescales, orders of magnitude larger than the usual Eddington limit.

\item The disk rapidly builds the BH from IMBH to SMBH masses, reaching $\gtrsim 2\times10^{6}\,\rm M_\odot$ in $\sim 1$\,Myr. Moreover if pairing has occurred, the inner BH binary appears to undergo the expected process whereby the secondary accretes more rapidly until the two BHs are brought to nearly equal mass, and they appear to grow at nearly identical rates, promoting near equal-mass BH-BH mergers.

\end{enumerate}

This scenario therefore provides a natural explanation for the origins and rapid formation of high-redshift quasars in the early Universe. It also has critical implications for mergers of supermassive BHs in the LISA bands, and if the processes above can operate at even higher mass scales ($\sim 10^{9}\rm M_\odot$), it could also provide an explanation for the seemingly-high rate of inferred SMBH mergers from pulsar timing \citep{AgazieAnumarlapudiArchibald_2023ApJ...951L...8A}. We also note that we see a hyper-magnetized, flux-frozen disk akin to those in \citet{HopkinsSquireSu_2024OJAp....7E..19H} but under very different physical conditions and with distinct numerical methods and resolution and detailed physics treatments (as did e.g.\ \citealt{GaburovJohansenLevin_2012ApJ...758..103G}) which suggests these are robust and can indeed power super-Eddington mass accretion rates. It also demonstrates that one can quickly reach true supermassive BH masses without invoking exotic channels such as ``direct collapse'' \citep{VolonteriRees_2005ApJ...633..624V,NatarajanTreister_2009MNRAS.393..838N,MayerKazantzidisEscala_2010Natur.466.1082M} of primordial gas clouds without fragmentation (indeed, here, fragmentation plays a key role {\it promoting} rapid BH growth) or other ``new physics'' (\S~\ref{sec:intro}).

In followup work we hope to make more detailed observational predictions. But observations will be challenging, since we predict this occurs on short time and small spatial scales in high-redshift systems, with comparable stellar and BH accretion bolometric luminosity, and the medium is still dusty and obscured at these times owing to the high gas densities. Still it is immediately striking that the bolometric luminosities ($\sim 5-50 \times 10^{44}\rm erg\,s^{-1}$, depending on the BH radiative efficiency), sizes ($\sim 10-100\,$pc), broadly comparable mix of stellar-and-AGN luminosity, and colors (given the predicted dust column densities at large radii from Fig.~\ref{fig:disk}) this simulation would exhibit during its later evolution ($\gtrsim 0.7\,$Myr, after coalescence and disk formation) are all remarkably consistent with the observed properties of the unexpected population of ``little red dots'' seen by JWST at redshifts $z \sim 4-10$ \citep[compare e.g.][]{KokorevCaputiGreene_2024ApJ...968...38K,AndikaJahnkeOnoue_2024A&A...685A..25A,KocevskiFinkelsteinBarro_2024arXiv240403576K}. 

Interestingly, the BHs here end up well {\it above} the $M_{\rm BH}-M_{\ast}$ (bulge mass), but well {\it below} the $M_{\rm BH}-\sigma_{\ast}$ (velocity dispersion) correlations observed at $z=0$ \citep{KormendyHo_2013ARA&A..51..511K,McConnellMa_2013ApJ...764..184M}. This may therefore provide an explanation for apparently ``overmassive'' BHs observed at high redshift (and some low-redshift outliers; \citealt{SethvandenBoschMieske_2014Natur.513..398S,TrakhtenbrotUrryCivano_2015Sci...349..168T,WalshvandenBoschGebhardt_2016ApJ...817....2W,LiepoldQuennevilleMa_2020ApJ...891....4L}), but also implies typical systems must grow {\it both} in bulge mass and BH mass before becoming ``typical'' $z=0$ massive classical bulges (as indeed predicted for more classical feedback-regulated models of the evolution of these properties; see \citealt{HopkinsHernquistCox_2008ApJS..175..356H,HopkinsCoxKeres_2008ApJS..175..390H}). 

Of course, there are many caveats and limitations to the simulations here which merit further study. Our resolution is limited: future work with super-resolution approaches like those in \citet{Angles-AlcazarQuataertHopkins_2021ApJ...917...53A} would enable robust mapping of the behavior interior to the radius of influence, and connection to GRMHD simulations like \citet{KaazLiskaJacquemin-Ide_2023ApJ...955...72K} to horizon scales. Improving the star formation model with approaches like \citet{GrudicGuszejnovHopkins_2021MNRAS.506.2199G} would free us from the universal IMF assumption and allow for actual predictions of exotic IMFs (and their effects) in high-redshift systems. In order to consider a ``controlled'' experiment, we utilize idealized initial conditions with pre-existing gas and magnetic fields and BH seeds, which we could improve by following these systems in fully-cosmological simulations with self-consistent seed BH formation models. And we consider only one, simplified (and relatively ``weak'') model for ``feedback'' (in the form of jets, winds, and radiation) from BH accretion (motivated by the study in \citealt{ShiKremerHopkins_2024arXiv240512164S} and slim disk models in \citealt{WataraiFukueTakeuchi_2000PASJ...52..133W,MadauHaardtDotti_2014ApJ...784L..38M}). But given that the hyper-magnetized accretion disk here appears to be qualitatively, fundamentally distinct from classical $\alpha$-disks on which these feedback models are based, it is important to understand the horizon-scale (hence radiation, jet and wind) properties of such disks in order to better inform next-generation sub-grid feedback models.

\section*{Acknowledgements}
Support for PFH was provided by NSF Research Grants 20009234, 2108318, NASA grant 80NSSC18K0562, and a Simons Investigator Award. Support for KK was provided by NASA through the NASA Hubble Fellowship grant HST-HF2-51510 awarded by the Space Telescope Science Institute, which is operated by the Association of Universities for Research in Astronomy, Inc., for NASA, under contract NAS5-26555. Numerical calculations were run on NSF/TACC allocation AST21010 and NASA HEC SMD-16-7592.

%%%%%%%%%%%%%%%%%%%%%%%%%%%%%%%%%%%%%%%%%%%%%%%%%%
\section*{Data Availability}
Simulation data involved in this work are available upon reasonable request to the authors.


\bibliography{bib}
\bibliographystyle{aasjournal}



\end{document}